\documentclass{PoS}

\usepackage{amsmath,amssymb}

\title{Using Gravitational Wave Observations to Probe Quantum Gravity}

\ShortTitle{Low energy probes for quantum gravity}

\author{\speaker{Justine Tarrant}\\
        University of the Witwatersrand\\
        E-mail: \email{justine.tarrant@wits.ac.za}}
    
\author{{Sergio Colafrancesco}\\
	University of the Witwatersrand\\
	E-mail: \email{sergio.colafrancesco@wits.ac.za}}

\abstract{Electromagnetic radiation is known to be associated with certain gravitational waves events, i.e. the collision of binary neutron stars. Establishing this connection is non-trivial. However, if electromagnetic counterparts could be produced by directly converting gravitons into photons, then a simple smoking gun test exists linking the two events. This model uses the general and conversion mechanism discussed by Raffelt and Stodolsky. Furthermore, because this mechanism is generic to the symmetries of general relativity and the standard model and because it assumes a quantised gravitational field, we may probe both the existence of the graviton and the scale at which quantum gravity effects (and thus very high-energy phenomena) become relevant using simple low-energy experiments.}

\FullConference{5th Annual Conference on High Energy Astrophysics in Southern Africa\\
		4-6 October, 2017\\
		University of the Witwatersrand (Wits), South Africa}

\begin{document}

\section{Introduction}

Electromagnetic counterparts are known to be associated with the production of gravitational waves \cite{TheLIGOScientific:2017qsa} in some cases, i.e. binary neutron star mergers and possibly black hole-neutron star mergers. These counterparts are formed via, as yet, unknown astrophysical processes, see \cite{Ioka:2017nzl} for some suggestions. The merger of binary neutron stars produce radiation across the entire electromagnetic spectrum \cite{GBM:2017lvd} in the form of gamma-ray bursts and broadband afterglow when these gamma-rays interact with the interstellar medium \cite{Alexander:2017aly}. Associating the gamma-ray burst with the gravitational wave event is non-trivial. This is because localizing events in the sky is difficult due to the low accuracy inherent in gravitational astronomy. It took an international effort to pinpoint the source of the electromagnetic radiation and conclude that the event was in range of the neutron star pair. 

It would therefore be beneficial to find a simple method for localizing the gravitational wave event and it's connection to gamma-ray bursts. Here we propose such a simple model that may be used to explain the origin of the electromagnetic radiation accompanying the gravitational waves, with the advantage of being a smoking gun test requiring less observational effort. That is, we show that it is possible to draw conclusions about the corresponding electromagnetic radiation using only low-frequency observations. Plausible low frequency experiments include SKA-Low and a lunar based radio telescope. 

The notion that photons and low-mass/zero-mass bosons, such as the graviton, can mix has been well studied in the literature. This theoretical principle stems from the earlier treatment of axion-photon mixing \cite{Peccei:1977hh,Kaplan:1985dv}. Using the idea of `mixing', as described by Raffelt and Stodolsky \cite{Raffelt:1987im} allows gravitational waves consisting of gravitons to be directly converted into photons as they pass through an external magnetic field. Of course, this implies that the gravitational field is quantized. Therefore, this model also provides a potential way to probe quantum gravity and its associated energy scales. Using the characteristics of relatively well studied environments, such as the known external magnetic field, their size and a phenomenological energy scale $M$, we may compute the probability that gravitons are converted to photons in the weak mixing limit.

The goal is to then find the parameter $M$ through observation of the spectra associated to the gravitational wave event and its associated electromagnetic radiation due to conversion. Then we compute the conversion probability and compare to the phenomenological value for $M$ which in the case for graviton-photon mixing is just $M_{Pl}$, the Planck mass. We would therefore be able to place a bound on the energy scale through any deviation from the Planck mass. Furthermore, such an observation would determine the existence of the graviton and therefore that the gravitational field is quantized.

We show that the conversion probabilities using $M_{Pl}$ are large enough to ensure that a significant amount of energy results from the conversion of gravitons into photons. This is promising because these large energies ($10^{35}$-$10^{37}$ erg) yield fluxes of up to $10^{-13}$ erg cm$^{-2}$ s$^{-1}$ for $\nu \lesssim 100$ kHz. Thus, there exists the potential for extrapolation to observable fluxes within the operating band of the SKA-Low ($\nu > 50$ MHz) or a putative lunar-based array ($\nu > 1$ MHz). Future work will expand on whether these aforementioned experiments can provide a realistic radio window onto quantum gravity.


This paper is structured as follows: section 2 provides a brief overview of the theoretical treatment of the mixing of gravitons and photons; our results are then presented in Section 3, and we draw conclusions in Section 4.

\section{Theoretical Background}

Raffelt and Stodolsky present a mechanism for the direct conversion of gravitons into photons and vice versa. This conversion requires the presence of an external magnetic field supplying one virtual photon in order to satisfy symmetry constraints, thus conversion of real gravitons to real photons is one-to-one, as the second photon in this context is virtual. The Lagrangian density governing this interaction is given by \cite{Ejlli:2013gaa}

\begin{equation}
\mathcal{L} = \mathcal{L}_{EM} + \mathcal{L}_{EH} + \mathcal{L}_{QED},
\end{equation}

\noindent where
\begin{equation}
\mathcal{L}_{EM} = -\frac{1}{4}F_{\mu\nu}F^{\mu\nu},
\end{equation}

\begin{eqnarray}
\mathcal{L}_{EH} = -\frac{1}{4}&(\partial_{\mu} h_{\alpha\beta}\partial^{\mu}h^{\alpha\beta} - \partial_{\mu} h\partial^{\mu}h + 2\partial_{\mu} h_{\mu\nu}\partial^{\nu}h\nonumber\\ &-2\partial_{\mu} h_{\mu\nu}\partial^{\rho}h^{\rho}_{\nu} ) -\frac{k}{2}h_{\mu\nu}T^{\mu\nu}_{EM},
\end{eqnarray}

\begin{equation}
\mathcal{L}_{QED} =\frac{\alpha^2}{90m_e^4}\left[(F_{\mu\nu}F^{\mu\nu})^2 + \frac{7}{4}(\tilde{F}_{\mu\nu}F^{\mu\nu})^2\right],
\end{equation}

\noindent where $\mathcal{L}_{EM}$ is the Lagrangian density of the electromagnetic field, $\mathcal{L}_{EH}$ is the linearised Einstein-Hilbert Lagrangian and finally, $\mathcal{L}_{QED}$ is the Lagrangian density accounting for QED effects due to vacuum polarization, also called the Euler-Heisenberg Lagrangian. The electromagnetic field tensor is given by $F_{\mu\nu}$, and $\tilde{F}_{\mu\nu} = \frac{1}{2}\epsilon_{\mu\nu\rho\sigma}F^{\rho\sigma}$ is its dual and $h_{\alpha\beta}$ is the graviton field. The electron mass is given by $m_e$ and $\alpha \approx 1/137$ is the fine structure constant.

\noindent From the Euler-Lagrange equations one obtains the equations of motion \cite{Raffelt:1987im}

\begin{equation}
\left[\omega + \left[\begin{matrix}
\Delta_{\perp} & \Delta_M & 0 & 0\\
\Delta_M & 0 & 0 & 0\\
0 & 0 & \Delta_{\parallel} & \Delta_M\\
0 & 0 & \Delta_M & 0
\end{matrix}\right] - i\partial_z\right]\left[\begin{matrix}
A_{\perp}\\ G_+\\ A_{\parallel}\\ G_{\times}
\end{matrix}\right] = 0
\end{equation}

\noindent where $\Delta_{\parallel}$, $\Delta_{\perp}$ are the momentum differences of the respective modes (polarizations) compared to photons of the same energy. The off-diagonal components are given by $\Delta_M = (B_e/2M)\sin\Theta$ with $\Theta$ the angle between the external field direction and the photon momentum. The energy scale $M$ represents a phenomenological parameter indicating interaction strength \cite{Raffelt:1987im}. The strength of the mixing is proportional to the ratio of the off-diagonal to the difference in diagonal terms:

\begin{equation}
\frac{1}{2}\tan 2\theta = \frac{\Delta_M}{\Delta_{\perp} - \Delta_{\parallel}}
\end{equation}

\noindent where $\theta$ is the mixing angle. The general solution to the above equations of motion, using the mixing angle $\theta$, is given by

\begin{equation}
\left[\begin{matrix}
A_{\perp}\\ G_+\\ A_{\parallel}\\ G_{\times}
\end{matrix}\right] = \mathcal{M}(z)\left[\begin{matrix}
A_{\perp}(0)\\ G_+(0)\\ A_{\parallel}(0)\\ G_{\times}(0)
\end{matrix}\right]
\end{equation}

\noindent where

\begin{equation}\label{eq2}
\mathcal{M}_{1,2}(z) = \left[\begin{matrix}
\cos\theta & -\sin\theta\\
\sin\theta & \cos\theta
\end{matrix}\right]\left[\begin{matrix}\Delta_{1,2} & \Delta_M\\
\Delta_M &0 \\
\end{matrix}\right]\left[\begin{matrix}
\cos\theta & -\sin\theta\\
\sin\theta & \cos\theta
\end{matrix}\right]^{-1}
\end{equation}

\noindent is the mixing matrix and $\{1,2\}$ run over $\{\perp,\parallel\}$. Here we consider only the well known weak conversion/mixing limit, in which the photon or graviton energy is well below the critical energy, i.e. the energy for which the mixing becomes significant. This also implies that the characteristic oscillation length, from graviton to photon and back, is much longer than the length scale considered, implying that $\theta << 1$

Supposing that gravitational waves are made up of gravitons, we may apply this mechanism to produce direct electromagnetic counterparts from the gravitational waves themselves via this conversion process. The fraction of gravitational radiation converted into electromagnetic radiation depends, in the weak mixing limit, only on the external magnetic field $B_e$, the size of the source $R$ and the graviton coupling $g = 1/M$ where $M = M_{Pl}$ is the Planck mass in this case as in \cite{Raffelt:1987im,Horns:2012kw}

\begin{equation}\label{eq1}
f \propto B_e^2 R^2 g^2.
\end{equation}

\noindent Thus larger field environments allow for further travel of gravitons and therefore allow more conversions to take place. We use (\ref{eq1}) in the computations that follow.

This conversion fraction is also given by $f=E_{EM}/E_{GW}$ so that the total energy released may be written as $E_{TOT} = (1+f)E_{GW} = E_{GW} + E_{EM}$. Obtaining an $f$ by experimental means then implies knowing the relationship between the spectra for the gravitational waves and the corresponding electromagnetic radiation. That is, by what fraction is the flux smaller in the electromagnetic radiation as compared with the gravitational wave spectrum.

\section{Results and Discussion}

\begin{table}[ht!]
	\begin{tabular}{ |p{3cm}||p{2cm}|p{2cm}|p{2cm}|p{2cm}|p{2cm}|  }
	\hline
	\multicolumn{6}{|c|}{Characteristic Environments} \\
	\hline
	Environment & $B_e$ (G) & $R$ (kpc) & f & $E_{\gamma}$ (erg) & F (erg.cm$^{-2}$s$^{-1}$)\\
	\hline
	\hline
	Pulsar & $10^{12}$ & $10^{-16}$ & $10^{-16}$ & $~ 10^{35}$ & $10^{-15}$\\
	\hline
	White Dwarf(Sirius B) & $10^{8}$ & $10^{-13}$ & $10^{-18}$ & $~ 10^{33}$ & $10^{-17}$\\
	\hline
	Starburst Galaxy(M82) & $10^{-5}$ & $1$ & $10^{-18}$ & $~ 10^{33}$ & $10^{-17}$\\
	\hline
	ISM & $10^{-9}$ & $10^{5}$ & $10^{-16}$ & $~ 10^{35}$ & $10^{-15}$ \\
	\hline
	Coma Cluster & $10^{-6}$ & $10^{3}$ & $10^{-14}$ & $~ 10^{37}$ & $10^{-13}$\\
	\hline
	Radio Galaxy(M87) & $10^{-5}$ & $10^{2}$ & $10^{-14}$ & $~ 10^{37}$ & $10^{-13}$\\
	\hline
\end{tabular}
\caption{Order of magnitude estimates for conversion probability $f$ and the flux $F$ for specific well characterized environments. $B_e$ is the external magnetic field for the given environment and $R$ is the size of the environment. The energy released as gravitational waves was $E_{GW} = 10^{51}$ ergs. The data was sourced from the following references: \cite{Raffelt:1987im} for the pulsar magnetosphere and the white dwarf, \cite{0034-4885-57-4-001,1538-4357-514-2-L79} for the ISM, \cite{bonafede} for Coma, \cite{Machalski:2008ir} for the Radio Galaxy and \cite{1992A&A...256...10R} for the starburst galaxy. \label{maintab}}
\end{table}

Table~\ref{maintab} provides order of magnitude estimates for the conversion probability and the resulting electromagnetic flux given an external magnetic field $B_e$, a characteristic size $R$. These values are calculated within the weak mixing limit. Using these relatively well characterized environments will allow one to infer an energy scale for quantum gravity effects. We will now discuss this in some detail.

The conversion probabilities in Table 1 are calculated using the phenomenological value for the graviton coupling $g =1/M_{Pl}$ where $M_{Pl}$ is the Planck mass, roughly $10^{19}$ GeV. The total flux $F$ for each event is calculated using $t = 10^{-3}$s, the time in which the energy is emitted in the form of gravitational waves \cite{Kawamura:2003hu}, and the fiducial distance to the gravitational wave source $d = 100$ Mpc, as well as the conversion fraction. Spectral studies will be considered in future work. The total energy released in gravitational waves, considered here, is roughly $10^{51}$ ergs \cite{Kawamura:2003hu}. The energy released as electromagnetic radiation is then $E_{\gamma} = fE_{GW}$.


One can see from Table 1 that the energy released through the conversion of gravitons to photons using the mechanism detailed by Raffelt and Stodolsky remains significant, similar in magnitude to solar flares occurring on the surface of the sun \cite{solarflares}. Furthermore the fluxes are quite bright and therefore potentially detectable. Since the gravitational wave spectrum appears over low frequencies, the resulting electromagnetic spectrum, which is directly related here (through the conversion), will appear over the low frequency band as well. The large magnitude of the fluxes at frequencies below 100 kHz means that the potential extrapolation of the spectrum to higher energies could produce potentially observable signals in current or future low-frequency experiments. In particular, SKA-Low or a hypothetical lunar based telescope would be possible candidates to detect events in which gravitational radiation is directly converted into photons. This will be investigated in future work.

We require an electromagnetic emission that comes directly from conversion of gravitons into photons. That is, there should not be another astrophysical explanation for the electromagnetic radiation. Provided that a candidate event, i.e. a binary neutron star merger, is observed, resulting in a electromagnetic spectrum, we may compare this spectrum to the gravitational wave spectrum coming from the corresponding gravitational wave event. Comparing the spectra will provide us with a fraction which is equal to the conversion probability $f$, i.e. we calculate the fractional change in energy from the gravitational waves to electromagnetic energy. This is simplified since the spectra will both have the same shape, in other words the distribution of photons will be the same, but having a lower flux. From this value for $f$ we may infer a value for the graviton coupling $g$ using (\ref{eq1}). This mechanism then provides a unique and simple way for probing the energy scales for quantum gravity. Using these spectra, we are able to place bounds on possible values for $g$ which may deviate from $1/M_{Pl}$.

\section{Conclusion}

Provided the weak mixing limit holds, we may use Eq.~(\ref{eq1}) to calculate $f$. We did this using the phenomenological value of $g = 1/M_{Pl}$. This gives us a reference when comparing to observation. The environments considered herein are relatively well characterized and produce conversion probabilities that are large enough to ensure a large total flux is obtained (up to $F \sim 10^{-13}$ erg cm$^{-2}$ s$^{-1}$ for $\nu \lesssim 10^2$ kHz). Studies will be conducted in future to determine whether the resulting spectrum may extrapolate to the observation bands of the SKA-Low ($\nu > 50$ MHz) or a future lunar-based array ($\nu > 1$ MHz).

This approach to producing electromagnetic counterparts, via the direct conversion of gravitons into photons, illuminates possible avenues for studying the existence of the graviton and its energy scale using simple low-energy experiments such as lunar based radio observatories and possibly even SKA-Low.


\section{Acknowledgments}

SC acknowledges support by the South African Research Chairs Initiative
of the Department of Science and Technology and National
Research Foundation, as well as the Square Kilometre Array (SKA).
This work is based on the research supported by the South African
Research Chairs Initiative of the Department of Science and Technology
and National Research Foundation of South Africa (Grant
No 77948). JT acknowledges support from the DST/NRF
SKA post-graduate bursary initiative.

\bibliographystyle{unsrt}
\bibliography{bib}

\end{document}